      [1999/12/01 v1.4c Il Nuovo Cimento]
\documentclass{cimento}

\usepackage{graphicx}  
\title{Evolution and nucleosynthesis in low mass Asymptotic Giant Branch stars}
\shorttitle{Evolution and nucleosynthesis in low mass AGB stars}
\author{S.~Cristallo\from{ins:x}}
\instlist{\inst{ins:x} Osservatorio Astronomico di Teramo (INAF) -
via Maggini, 64100, Teramo (Italy)}
\PACSes{\PACSit{97.10.Cv}{Stellar structure, interiors, evolution,
nucleosynthesis, ages} \PACSit{97.10.Tk}{Abundances, chemical
composition} }
\begin{document}

\maketitle

\begin{abstract}
People usually smile when astrophysicists assert that we are {\it
sons of the stars}, but human life confirms this sentence: about
65\% of the mass of our body is made up of oxygen, carbon occurs
in all organic life and is the basis of organic chemistry,
nitrogen is an essential part of amino acids and nucleic acids,
calcium is a major component of our bones. Moreover, phosphorus
plays a major role in biological molecules such as DNA and RNA
(where the chemical codes of life is written) and our blood
carries oxygen to tissues by means of the hemoglobin (an iron
pigment of red blood cells). All these elements have been created
in stars. I just list some examples related to human body, but
also common element such as aluminum, nickel, gold, silver and
lead come from a pristine generation of stars. The abundances in
the Solar System are in fact due to the mixing of material ejected
from stars that polluted the Universe in different epochs before
the Sun formation, occurred about 5 billion years ago, after the
gravitational contraction of the proto-solar cloud. Low mass AGB
stars (1$<$$M$/M$_{\odot}$$<$3) are among the most important
polluters of the Milky Way, because of the strong winds eroding
their chemically enriched envelopes. They are responsible for the
nucleosynthesis of the main component of the cosmic s-elements.
\end{abstract}

\section{Introduction}
The majority of the isotopes heavier than iron (A$\ge$56) are
synthesized by neutron capture processes. The observed heavy
elements distribution shows the presence of two main components,
correlated to different nucleosynthetic processes: the s (slow)
process  and the r (rapid) process (on the basis of the
definitions given by \cite{b2hf} in their pioneering work). The r
process requires high neutron densities, and it is believed to
occur during explosive phases of stellar evolution (Novae,
SuperNovae and/or X-rays binaries). The s process is characterized
by a slow neutron capture with respect to the corresponding
$\beta$ decay: stable isotopes capture neutrons, while the
radioactive ones decay ($\beta^-$ or $\beta^+$) or capture a free
electron. These isotopes are mainly created in the Thermally
Pulsing AGB (TP-AGB) phase of low mass stars (1 $\leq$
M/M$_{\odot}$ $<$ 4), where freshly synthesized elements are
carried out to the surface by means of a recurrent mechanism
called Third Dredge Up (TDU) (see \cite{stra06} and references
therein).
\begin{figure}[tb]
\centering
\includegraphics[width=8.5cm]{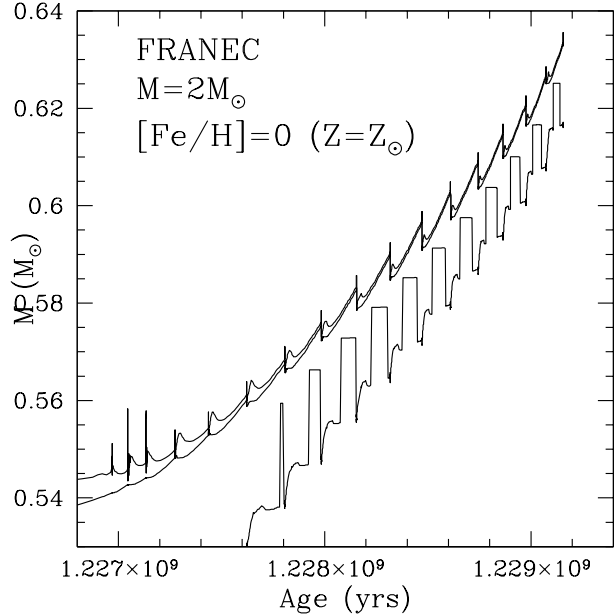}
\caption{Evolution of the positions in mass of the inner border of
(top to bottom): convective envelope, H-burning shell and most
energetic mesh of the He-burning shell, during the Thermally
Pulsing AGB phase of a model with initial mass M=2M$_\odot$ and
solar metallicity..} \label{mcore}
\end{figure}
In this phase the stellar structure consists of a partially
degenerate carbon-oxygen core, an He shell separated from an H
shell by the He-intershell region and by a convective envelope
(see Fig.\ref{mcore}). The energy required to supply the surface
irradiation is mainly provided by the H burning shell, located
just below the inner border of the convective envelope. This
situation is recurrently interrupted by the growing up of
thermonuclear runaways, driven by violent He-burning ignitions. As
a consequence of a Thermal Pulse (TP), the region between the two
shells (He-intershell) becomes unstable against convection (for a
short period), the external layers expand and, later on, the H
shell burning temporarily dies down. In the He-intershell, He is
partially converted into carbon. During the AGB phase, main
neutron sources are the $^{13}$C($\alpha$,n)$^{16}$O reaction,
active in radiative layers during the interpulse period
\cite{stra95}, and the $^{22}$Ne($\alpha$,n)$^{25}$Mg reaction,
marginally activated within the convective shell originated by the
TP. In order to obtain a sufficient amount of $^{13}$C for the
activation of the s-process, a diffusion of protons from the
H-rich envelope into the $^{12}$C-rich radiative zone is needed:
the diffused proton are captured from the abundant carbon via the
$^{12}$C(p,$\gamma$)$^{13}$N($\beta^-$)$^{13}$C nuclear chain,
leading to the formation of a tiny $^{13}$C-pocket.

\section{The $^{13}$C pocket and the nuclear network}

\begin{figure}[tb]
\centering
\includegraphics[width=8.5cm]{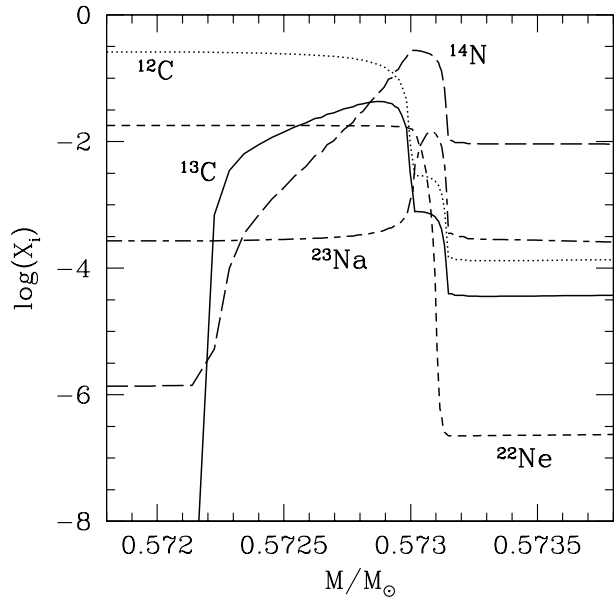}
\caption{Abundances profiles in the $^{13}$C pocket formed after
the 3$^{th}$ TDU episode ($M=2M_\odot$ model with solar
metallicity).} \label{tsol}
\end{figure}

A major improvement in our stellar evolution code is the
introduction of a physical algorithm for the treatment of the
convective/radiative interface at the inner border of the
convective envelope \cite{cri01,stra06}. During TDU episodes, the
opacity of the envelope (H-rich) is significantly larger than the
opacity of the underlying H-exhausted (and He-rich) region. This
fact causes an abrupt change of the temperature gradient at the
inner border of the penetrating convective envelope. In this
condition, the convective boundary becomes unstable, because any
perturbation causing an excess of mixing immediately leads to an
increase of the opacity and, in turn, to an increase of the
temperature gradient. However, the steep pressure gradient that
develops immediately below the formal border of the convective
envelope limits the penetration of the instability, so that the
average convective velocity should rapidly drop to zero. In order
to mimic this behavior, we assume that in the region underlying
the formal convective boundary, the average velocity follows an
exponential decline, namely

\begin{equation} \label{param}
v=v_{bce}\exp{ \left( -\frac{d}{\beta H_P} \right) } \;     ,
\end{equation}
where {\it d} is the distance from the formal convective boundary,
$v_{bce}$ is the velocity of the most internal convective mesh,
$H_P$ is the pressure scale height at the formal border of the
convective envelope (defined by the Schwarzschild criterion) and
$\beta$ is a free parameter. The partial diffusion of protons in
the top layers of the He shell gives naturally rise to the
subsequent formation of a $^{13}$C-rich tiny layer, the so-called
$^{13}$C pocket. In Fig. \ref{tsol} we plot the chemical profiles
of trace isotopes in the region around the $^{13}$C pocket.
The $^{13}$C pocket (solid line) partially overlaps with a more
external $^{14}$N pocket (long-dashed line). The maximum of the
$^{14}$N coincides with the region where the protons diffused from
the convective envelope at the epoch of the TDU were abundant
enough to allow a full CN cycle. This is also the region where the
$^{23}$Na (short-dashed-long-dashed line) has been efficiently
produced by the $^{22}$Ne(p,$\gamma$)$^{23}$Na reaction.  Indeed,
at the end of the previous TP, the mass fraction of $^{22}$Ne
(short-dashed line) in the top layer of the He-rich intershell is
of the order of 2$\times 10^{-2}$ and, around the $^{14}$N
maximum, this $^{22}$Ne is almost completely converted into
$^{23}$Na, leading to the formation of a tiny $^{23}$Na pocket.\\
The introduction of the exponentially decaying profile of
convective velocities automatically implies an assumption on the
value of the free parameter $\beta$: in the models presented in
this work, we assume a value $\beta$=0.1. In order to justify such
a choice we perform some tests on a star with initial mass
$M=2M_\odot$ and different metallicities ($Z$=Z$_\odot$ and
Z=$1.0\times 10^{-4}$). We calculate the same sequences
TP-Interpulse-TP, starting from the first TP followed by TDU, and
we evaluate the consequences of changing this parameter in the
range 0$< \beta <$0.2 (Cristallo et al., {\it in preparation}).
While the amount of material interested by a single TDU episode
increases by more than a factor of 3 from the case with no
velocity profile ($\beta$=0) and the extreme case ($\beta$=0.2),
the $^{13}$C$_{eff}$, available for the heavy elements
nucleosynthesis, grows up with increasing $\beta$, reaches a
maximum in correspondence of $\beta$=0.1 and then decreases down
to negative values. We stress the fact that the net $^{13}$C that
contributes to the s-process nucleosynthesis is represented by
$^{13}$C$_{eff}$, defined as the difference between the $^{13}$C
and the $^{14}$N mass fractions ($^{14}$N is in fact a strong
neutron poison via the $^{14}$N(n,p)$^{14}$C reaction). Our result
derives from a combination of two different physical processes:
the growing of the $^{13}$C$_{eff}$ is a direct consequence of the
greater efficiency of the velocity profile algorithm, while its
decreasing for large $\beta$ values is due to the fact that the
convection efficiency in the zone interested by the velocity
profile starts to be too high. This leads the most external layers
of the He-intershell to be fully mixed and to consequently show an
envelope-like H abundance, instead of the H profile needed for the
formation of the $^{13}$C pocket. We therefore propose an
efficient way to calibrate the free parameter affecting the
velocity profile algorithm, confining its possible values in a
narrow range, and we adjust it in order to obtain the maximum
$^{13}$C pocket expected for the s-process nucleosynthesis. \\
The simultaneous solution of the stellar structure equations and a
full network including all the relevant isotopes up to the
termination point of the s-process path (Pb-Bi) requires a
relevant computational power. For this reason, a post-process
nucleosynthesis calculation, based on AGB stellar models computed
with a restricted nuclear network, was generally preferred
\cite{ga98}. The coupling of a stellar code with a complete
nuclear network has not been feasible so far, but this limitation
has been overcome thanks to last generation of computers and to
the adoption of smart algorithms to invert sparse matrixes of huge
size. AGB models presented in this work are therefore calculated
by using an extended set of nuclear processes including all
chemical species involved in the s-process nucleosynthesis
\cite{cri04}. The models presented here have been obtained by
including into the FRANEC stellar evolution code
\cite{chi98,stra06} a full nuclear network, containing about 500
isotopes (from H to the Pb-Bi-Po ending point) linked by more than
750 reactions. Reaction rates of isotopes involving charged
particles are generally taken from the NACRE compilation
\cite{angulo}, while the neutron captures cross sections are
mainly taken from \cite{bao}. Weak-interaction rates (electron
captures, and $\beta$ decays) are interpolated as a function of
the temperature and electron density, while at temperatures lower
than $10^6$~K we assume a constant value equal to the terrestrial
one. The presence of isomeric states of unthermalized isotopes,
which lead to ramifications of the s-process flux, is followed in
detail. This network is continuously upgraded according to
the latest theoretical and experimental nuclear physics improvements. \\

\section{Low temperature C-enhanced opacities}\label{opa}

The FRANEC models benefit of another important improvement, whose
effects are particularly important at low metallicities: the
introduction of C-enhanced low temperature opacities. In
principle, a stellar evolutionary code should account for the
variations of the envelope chemical composition; in practice, only
variations of the main constituents (H and He) are usually
considered. Nevertheless, in the atmosphere of an AGB star, both C
dredge-up and the possible conversion of C into N (by CN cycle)
substantially affect the molecular contribution to the opacity for
temperatures lower than 4000 K. This is due to the fact that, when
a carbon excess is present (i.e. when the number ratio C/O$>$1),
carbon bearing molecules  (C$_2$, CN, C$_2$H$_2$, C$_3$) start
forming and their contribute to opacity become dominant.
Therefore, in order to overcome this limitation, we calculate low
temperature C- and N-enhanced opacity tables, suitable for AGB
model calculations \cite{cri07}. The new AGB models succeed in
reproducing the photometric and spectroscopic properties of their
observational counterparts (see Section \ref{lowz}).
\begin{figure}[tb]
\begin{center}
\includegraphics[width=8.5cm]{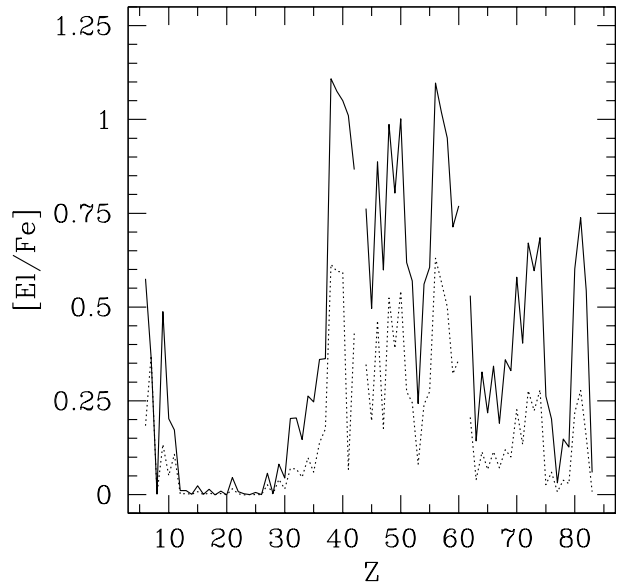}
\end{center}
\caption{Surface composition of a model with M=2M$_\odot$ and
solar metallicity after the fifth TDU (dotted curve) and the last
TDU (solid curve).} \label{disol}
\end{figure}

\section{Solar metallicity model}

The final C/O envelope ratio in the solar metallicity model is
1.88, while it becomes larger than 1 when the core mass is about
0.6 $M_\odot$ and the bolometric magnitude is about $-$5.0 mag, in
good agreement with the Galactic C-star luminosity function
\cite{guenda}. A $^{13}$C pocket, whose mass extension decreases
pulse after pulse, forms after each TDU episode. The $^{13}$C in
the first pocket is only partially burnt during the interpulse and
the residual is engulfed into the convective zone generated by the
subsequent TP. In this case, the $^{13}$C burning takes place at
the bottom of the convective shell ($T\sim 1.6\times 10^8$ K)
producing a very high neutron density, this fact implying
interesting peculiarities on the following nucleosynthesis (see
\cite{cri06}). In particular, some branchings, which remain closed
during a standard radiative $^{13}$C burning and are marginally
activated during the $^{22}$Ne($\alpha$,n)$^{25}$Mg burning, are
now open: in a very short temporal step ($\Delta T<3$ years) we
obtain a consistent production of neutron-rich isotopes normally
by-passed by the standard radiative $^{13}$C s-process, among
which $^{60}$Fe, $^{86}$Kr, $^{87}$Rb and $^{96}$Zr. Concerning
the standard radiative s-process nucleosynthesis, all the elements
\begin{figure}[tb]
\begin{center}
\includegraphics[width=12cm]{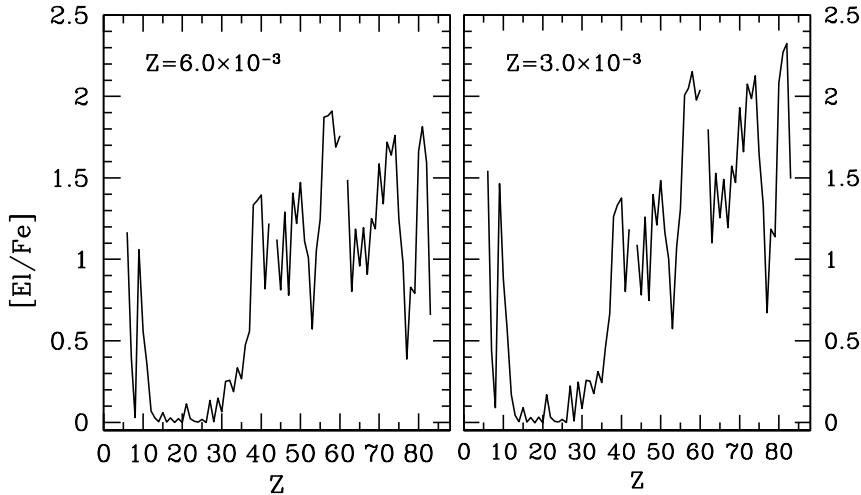}
\end{center}
\caption{Surface compositions of models with M=2M$_\odot$ and
Z=$6.0\times 10^{-3}$ (left Panel) and Z=$3.0\times 10^{-3}$
(right Panel).} \label{di_int}
\end{figure}
(from Sr to Pb) result enhanced. We find that the abundance of Sr,
Y and Zr at the first s peak, the so-called ls elements ({\it
light} s elements), is comparable with the one of the hs elements
({\it heavy} s elements) Ba, La, Ce, Pr, Nd at the second s peak
(see Fig. \ref{disol}). Lead is under-produced with respect
barium. The [hs/ls] \footnote{We use the standard spectroscopic
notation [A/B]=log($N$(A)/$N$(B))$-$log($N$(A)/$N$(B))$_\odot$.}
attained in the envelope when C/O=1 is in good agreement with
those measured in galactic C(N) giants \cite{ab02}.

\section{Intermediate metallicity models}

The $^{13}$C($\alpha$,n)$^{16}$O reaction (the main source of
neutrons) is primary-like, i.e. not directly affected by the
metallicity of the pristine material. However, since the build-up
of heavy nuclei requires neutron captures starting from Fe seeds,
the s-process is expected to decline with metallicity (it is
therefore of secondary nature). Nevertheless, while the iron seeds
scale with the metallicity, when decreasing the number of iron
seeds the number of neutrons available per seed is larger. The
dependence of s-process yields at different metallicities is
therefore very complex and non linear (see e.g. \cite{tra99}). At
metallicities Z$\sim(3\div 6)\times 10^{-3}$ the s-process
distribution is peaked around the hs elements (see Fig.
\ref{di_int}). Moreover, we obtain an increase in the relative
production of carbon and fluorine with decreasing the metallicity:
this is a direct consequences of the greater efficiency of the TDU
episodes, when the convective envelope penetrates into the
$^{12}$C-rich He-intershell (see \cite{stra03}). We want to stress
the fact that the production of $^{12}$C is of primary nature (it
is in fact produced by the activation of the triple $\alpha$
reaction in the He-intershell). The nature of the fluorine
production, that basically depends on the nucleosynthesis of
$^{15}$N (see e.g.
\begin{figure}[tb]
\begin{center}
\includegraphics[width=12cm]{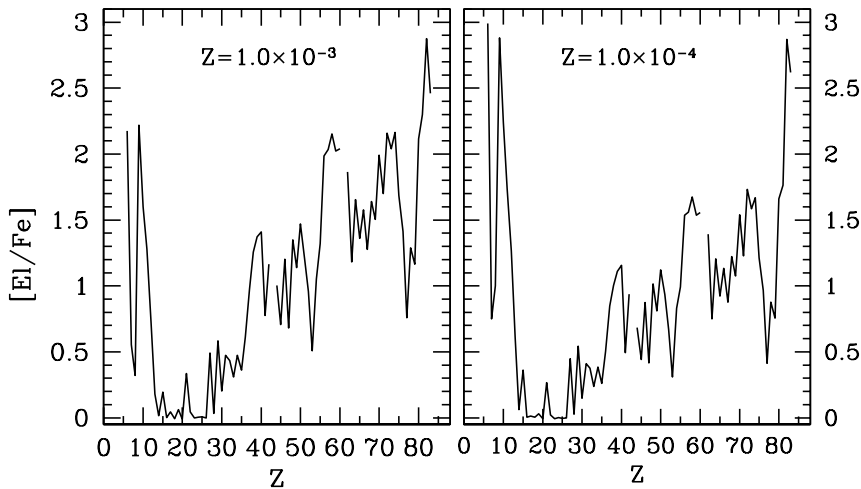}
\end{center}
\caption{Surface compositions of models with M=2M$_\odot$ and
Z=$1.0\times 10^{-3}$ (left Panel) and Z=$1.0\times 10^{-4}$
(right Panel).} \label{di_low}
\end{figure}
\cite{lu04}) is more complex, being partly of primary nature (due
to the contribution of the $^{13}$C pocket) and partly of
secondary nature (due to the contribution of the ashes of the
H-burning shell). Finally, let us note that at these metallicities
the production of lead (mainly $^{208}$Pb) starts being very
efficient.

\section{Low metallicity models} \label{lowz}

\begin{figure}[tb]
 \begin{center}
 \includegraphics[width=8.5cm]{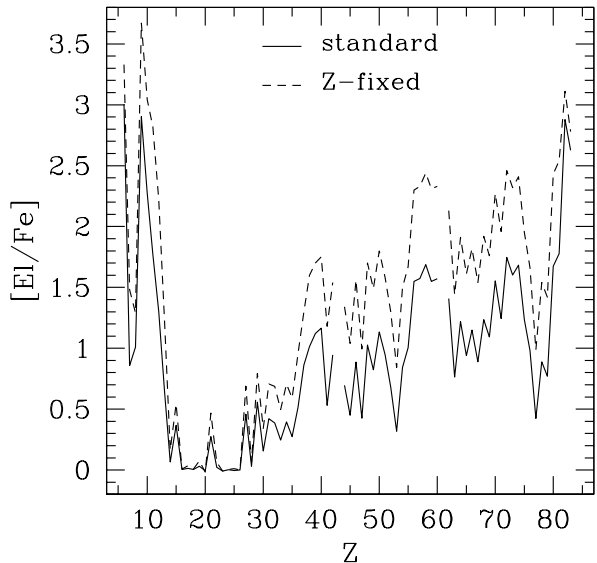}
  \caption{Surface compositions of a model with M=2M$_\odot$ and
Z=$1.0\times 10^{-4}$. The solid line refers to our {\it standard}
model; the dashed line refers to a model with the same initial
inputs but with a different opacity treatment ({\it Z-fixed}
model). See text for details.}
  \label{compall}
 \end{center}
\end{figure}

At low metallicities, most of the seeds are converted into
$^{208}$Pb, at the termination point of the s-process fluency (see
e.g. \cite{dela}). Trends illustrated in Fig. \ref{di_low} (which
relates to the $Z=1\times$10$^{-3}$ and to the
$Z=1\times$10$^{-4}$ models) confirm the previous sentence: a
consistent lead production ($Z$=82) is in fact found in both
models. As already stressed in the previous Section, the lower the
metallicity is, the larger surface carbon overabundances are
found: [C/Fe]$\sim$2.2 at $Z$=1$\times$10$^{-3}$ and
[C/Fe]$\sim$3.0 at $Z$=1$\times$10$^{-4}$. \\ Low metallicity
stars enriched in s-process elements show consistent enhancements
in the hs region, spanning in the range 0.8$<$[hs/Fe]$<$2.3~. Even
if our final [hs/Fe] value lays within the observed range, we
cannot reproduce the observed spread with a single evolutionary
model. In spite of this, we want to point out the importance of
the opacity coefficients on the stellar evolution of low
metallicity models. As already outlined in Section \ref{opa}, in
our models the opacity change caused by the variation of the
internal chemical composition has been taken into account by
linearly interpolating between tables with different carbon,
nitrogen and helium abudances \cite{cri07}. However, the procedure
followed by the majority of modelers is to adopt the opacity
coefficients calculated at the initial metallicity and to use them
for the entire evolution. In order to evaluate the possible
effects induced by these different opacity treatments, we compute
another model at $Z$=1$\times$10$^{-4}$ by using for the entire
evolution the opacity coefficients calculated for the initial
metallicity (we refer to this calculation as the {\it Z-fixed}
case) and we compare it with respect to the $Z$=1$\times$10$^{-4}$
model already presented ({\it standard} case). The TP-AGB phase of
the {\it Z-fixed} case is longer (a factor of 2 more than the {\it
standard} model), this implying a natural increase in the number
of TPs followed by TDU (N=49 with respect to the 15 found in the
{\it standard} model). Consequently, the surface abundances of the
{\it Z-fixed} case are larger with respect to the {\it standard}
case (see Fig. \ref{compall}), showing a better agreement with
observations. We argue that these difference in the surface
distributions are due to the mass loss rate, which results
less efficient in the {\it Z-fixed} model with respect to the {\it
standard} case, due to the fact that the surface temperature is
always larger and the radius is less expanded \cite{criphd}. As a
consequence of so different mass-loss rate histories, in the {\it
Z-fixed} case we find a large increase of the total dredged up
material during the AGB phase with respect to the {\it standard}
case and, therefore, the resulting final heavy element
enhancements result larger. Our results seem therefore to suggest
that the model computed with opacity tables at fixed metallicity
is more efficient in reproducing spectroscopic data. There are
however observational counterparts supporting the opposite
conclusion, driven by the study of C-stars in Dwarf Spheroidal
Galaxies (DSGs); we refer in particular to the Draco Dwarf
Spheroidal Galaxy \cite{aa85,dom04}. The observed giants belonging
to this DSph present surface temperatures in the range
3.55$<$logT$<$3.61~. Note that this range is never attained by the
{\it Z-fixed} case, while it is entirely spanned by the {\it
standard} case (see Fig.4 in \cite{cri07}).\\ Another important
aspect which has to be addressed is the choice of the mass-loss
history. Actually, observation cannot lead modelers in
parameterizing the mass-loss at low metallicities, basically due
to the fact that up to date a clear determination of the
gas-to-dust ratio (needed to properly calculate the mass-loss) has
not been feasible at non-solar metallicities \cite{la07,mat07}.
For this reason we adopt the same mass-loss history at all
metallicities (see \cite{stra06} for a description of our choice).
Preliminary results show however that when using a different
(milder) mass-loss \cite{rei} the surface overabundances increase
on average for 0.2 deX (Cristallo et al. {\it in preparation}). \\
Another source of uncertainty in modelling AGB stars at low
metallicity is the choice of the mixing length parameter, which
has been calibrated by calculating a Standard Solar Model. If
using initial abundances from \cite{lo03}, we fit the observed
properties of the Sun by using a value of the mixing length
parameter $\alpha_{m.l.}$=2.1 (see \cite{pi07}). Due to the
current lack of indications for the assumption of a different
$\alpha_{m.l.}$ value at low metallicities, we adopt the same
value at all metallicities: preliminary results indicate however
that when using a lower value (i.e. $\alpha_{m.l.}$=1.8) the
surface overabundances decrease on average for 0.2 deX (Cristallo
et al, {\it in preparation}).

\section{Very low metallicity models}

In the case of very metal poor AGBs, we find that, for a given
metallicity, there exists a lower mass limit for which normal AGB
evolution may occur \cite{cri07}. Stars with smaller initial mass
experience protons engulfment episodes during the first fully
developed TP (see e.g. \cite{ho90,iwa04}), which lead to a
peculiar s-process nucleosynthesis, a low $^{12}$C/$^{13}$C and a
significant synthesis of primary N. Owing to the low entropy
barrier, protons are engulfed within the He-intershell, where the
high temperature and the large C abundance induce a violent
H-burning flash. The consequent synthesis of $^{13}$C allows the
activation of the $^{13}$C($\alpha$,n)$^{16}$O reactions and a
very large neutron density is attained. Thus, the production of
neutron-rich nuclei, normally by-passed by a standard s-process,
can occur in these stars. Later on, when the intershell cools
down, a deep TDU occurs. Since in the envelope the C/O becomes
grater than 1, the large N abundance favors the formation of CN
molecules, thus inducing a further increase of the radiative
opacities. This new set of very low metallicity models, that
represents the current frontier of AGB modelling, is still under
analysis.

\section{Conclusions}

The present work demonstrates that, nowadays, the computational
power allows the coupling between a stellar evolutionary code and
a full nuclear network. A different treatment of the internal
border of the convective envelope with respect to a bare
Schwarzschild criterion allows the formation of the so-called
$^{13}$C pocket. For the first time in the literature, it has been
possible to directly compare theoretical models and observational
data both from a physical (luminosities, surface temperatures) and
a chemical (light elements and heavy elements abundances) point of
view . Moreover, we furnish a uniform  set of yields (from
hydrogen to lead) at different metallicities. The importance of
the adopted mass loss rate and of molecules contribution to
opacity has been pointed out. While the first problem is still
under analysis, the second one has been solved by using opacity
tables with enhanced carbon and nitrogen abundances. Finally, we
have addressed the importance of very low metallicity AGB models,
by describing first preliminary (and promising) results.

\acknowledgments I sincerely thanks Prof. O. Straniero and Prof.
R. Gallino for their invaluable expertise and their deep knowledge
of the arguments treated in this paper. Without a continuous
interaction and ideas sharing with them I would not acquire the
needed scientific uprightness to make research in a proper way.

\end{document}